\documentclass[twocolumn]{aastex6}

\usepackage{geometry}                
\usepackage{graphicx}
\usepackage{amsmath,amssymb}
\usepackage{epstopdf}
\usepackage{url} 

\begin{document}

\title{3D Simulations and MLT: I.  Renzini's Critique}

\author{W. David Arnett\altaffilmark{1}  
}
\author{Casey Meakin\altaffilmark{1,2}
}
\author{Raphael Hirschi\altaffilmark{3,4}
}
\author{Andrea Cristini\altaffilmark{3,7}
}
\author{Cyril Georgy\altaffilmark{5}
}
\author{Simon Campbell\altaffilmark{6}
}
\author{Laura J. A. Scott\altaffilmark{3}
}
\author{Etienne A. Kaiser \altaffilmark{3}
}


\altaffiltext{1}{Steward Observatory, University of Arizona, 
933 N. Cherry Avenue, Tucson AZ 85721 
}
\altaffiltext{2}{Karagozian \& Case, Inc., Glendale, CA 91203
}
\altaffiltext{3}{Astrophysics Group, Keele University, Lennard-­Jones Laboratories, Keele, ST5 5BG, UK
}
\altaffiltext{4}{Kavli IPMU (WPI), The University of Tokyo, Kashiwa, Chiba 277­ 8583, Japan
}
\altaffiltext{5}{Department of Astronomy, University of Geneva, Ch. Maillettes 51, 1290 Versoix, Switzerland
}
\altaffiltext{6}{Monash Centre for Astrophysics, School of Physics and Astronomy,
Monash University, Clayton, Australia 3800
}
\altaffiltext{7}{Department of Physics and Astronomy, University of Oklahoma, Norman OK 73019
}


\email{Corresponding author: wdarnett@gmail.com}


\begin{abstract}
\cite{alvio87}  wrote an influential critique of mixing-length theory  \citep[][MLT]{bv58} as used in stellar evolution codes, and concluded that three-dimensional  (3D) fluid dynamical  simulations were needed to clarify several important issues. We have critically explored the limitations of the numerical methods and conclude that they are approaching the required accuracy \citep{321D,woodward2015}.
Implicit large eddy simulations (ILES) automatically connect large scale turbulence to a Kolmogorov cascade below the grid scale, allowing
turbulent boundary layers to remove singularities that appear in the theory.  Interactions between coherent structures give multi-modal behavior, driving intermittency and fluctuations.
Reynolds averaging (RA) allows us to abstract the essential features of this dynamical behavior of boundaries which are appropriate to stellar evolution,
and consider how they relate static boundary conditions (Richardson, Schwarzschild or Ledoux).  
We clarify several questions concerning when and why MLT works, and does not work, using both analytical theory and 3D high resolution numerical simulations.  
The composition gradients and boundary  layer structure which are produced by our simulations
suggest a self-consistent approach  to boundary layers, removing the need for ad hoc procedures for 'convective overshooting' and `semi-convection'.
In a companion paper we quantify the adequacy of our numerical resolution,  determine of the length scale of dissipation (the `mixing length') without astronomical calibration, quantify agreement with the four-fifths law of Kolmogorov for weak stratification, and
extend MLT to deal with strong stratification. 
\end{abstract}

\section{Introduction}
The standard treatment of convection in stellar evolution theory , \cite[``mixing-length theory'',][MLT]{bv58}) uses 
a semi-empirical, ``engineering" approach,  based upon an approximate conceptual model due to Prandtl \citep{kippen,ddc,hansenkawaler2}. It is local, requires calibration, and has little connection to modern methods used by the turbulence community \citep{davidson,pope}. The mathematical basis of MLT is the B\"ohm-Vitense cubic equation for adiabatic excess, which is {\em not unique}, but may be derived from more than one physical model (e.g., different patterns of assumed flow, \cite{bv58,am11turb,nathan2014,gabriel18}).  

Thermonuclear burning in stars gives composition change, which drives stellar evolution.
Convection is a fundamentally  important process because it (1) moves energy and (2) gives mixing of fuels and ashes. The complete problem is daunting  \citep{am16key,miro}; at present it is not possible to directly simulate turbulent convection in stars over evolutionary times.
This paper is an early step and simplifications are needed.
We first consider convection separately from the effects of rotation, binary companions, and magnetic fields, and consider only a sector of a convective shell. This is essentially the classic B\"ohm-Vitense problem, which is standard for stellar evolution.
We also consider turbulence, wave generation, and boundary physics, each of which were not part of the B\"ohm-Vitense problem, but should have been.  

In \S2 we summarize Renzini's critique, and  refer back to these specific problems in subsequent sections.
We use this critique as a foundation on which we illustrate the direct relevance of 3D simulations to stellar evolutionary practice. 
In \S3 we present our methods of simulation and analysis, which allow us to study flows of relatively strong turbulence (Reynolds numbers up to $\sim 10^4$). 
In \S4 we reject the ``blob'' picture of MLT, examine the length scale for turbulent damping, the balance between driving and damping, and introduce the time-dependent turbulent kinetic energy (TKE) equation.
In \S5  we present simulation results for the evolution of the TKE, including multi-modal behavior and wave generation.
In \S6 we discuss composition gradients and boundary layer structure.\footnote{We avoid the terminology  {\em shallow} for 
{\em weakly-stratified} convection, which can be confusing. Weak stratification tends to occur {\em deep} in the central regions of stars (which typically contain a few pressure scale heights), while convection at stellar surfaces is highly stratified (often $\sim 20$ pressure scale heights).
}
 In \S7 we summarize this paper.

\section{Renzini's critique}\label{s-alvio}
In his classic critique of MLT,  \cite{alvio87} focused on convective ``overshooting'', which denotes attempts to deal with boundaries in a merely local theory. In discussing some proposed overshooting algorithms, he identified several fundamental problems with MLT that he labeled as  ``embarrassing":
\begin{enumerate}
\item The {\em ends} problem: infinite accelerations and decelerations are required at the beginning and end of the MLT trajectory. MLT does not define a boundary, so that additional physics must be assumed, usually involving the Schwarzschild or the Ledoux linear stability condition.
\item The {\em two lengths} problem:  are the path length and the size of the ``blob" the same?
\item The {\em resolution} problem:  are lengths resolved which are smaller than the mixing length  $\ell$ (or a pressure scale height $ H_P$)? Is a convective cell of the order of the zone size?
\item The {\em fluctuations} problem: turbulent fluctuations are ignored even if not small, so that  MLT does not deal with a ``storm of the century" event, nor the accumulated effects of fluctuations.
\item The {\em origin} problem: what is the flow pattern near the center of the star? Conservation of baryons requires that any flow into the central regions must be balanced by a flow out; if the radial velocity is non-zero at the origin, then its gradient must be zero. 
\item The {\em braking} problem: what causes  the flow to turn, and be contained in the convection zone? This requires buoyancy braking, which is not in MLT; and is often patched by "overshoot" prescriptions.
\item The {\em dynamics} problem:  Renzini's ``wind and water line" problem, or ``boundary layer'' dynamics. How are waves generated? How do convective boundaries grow and recede? 
\item The {\em non-locality} problem: what are the turbulent trajectories, and do distant regions affect local motion?
\item The {\em flux of turbulent kinetic energy} problem: this is ignored in MLT but has been demonstrated to be non-negligible  for strongly-stratified flows (e.g., surface convection zones in stars).
\item The {\em composition} problem: composition is assumed to be homogeneous in MLT, which is violated in regions having  nuclear burning, for example.
\end{enumerate}

We will take these problems one by one,  illustrating the strengths and weaknesses of MLT, and the improvements suggested by  3D simulations and their theoretical analysis. It has become traditional to complain about the flaws of MLT, but as \cite{alvio87} emphasized, MLT works surprisingly well in some respects.
We focus on the questions: {\em why does MLT work at all?}  and {\em what is still missing?}

It is appropriate here to warn the reader that two assumptions of MLT are violated by 3D simulations: (1) the turbulent velocity field is nonlocal, and (2) the net radial turbulent kinetic energy flux is not always negligible. Renzini's ``embarrassments'' are affected by these differences, which modify physics at boundaries.

\section{Simulation methods}\label{Ssims}

\begin{figure*}[h]
\figurenum{1}
\label{fig1}
\includegraphics[angle=0,scale=0.45]{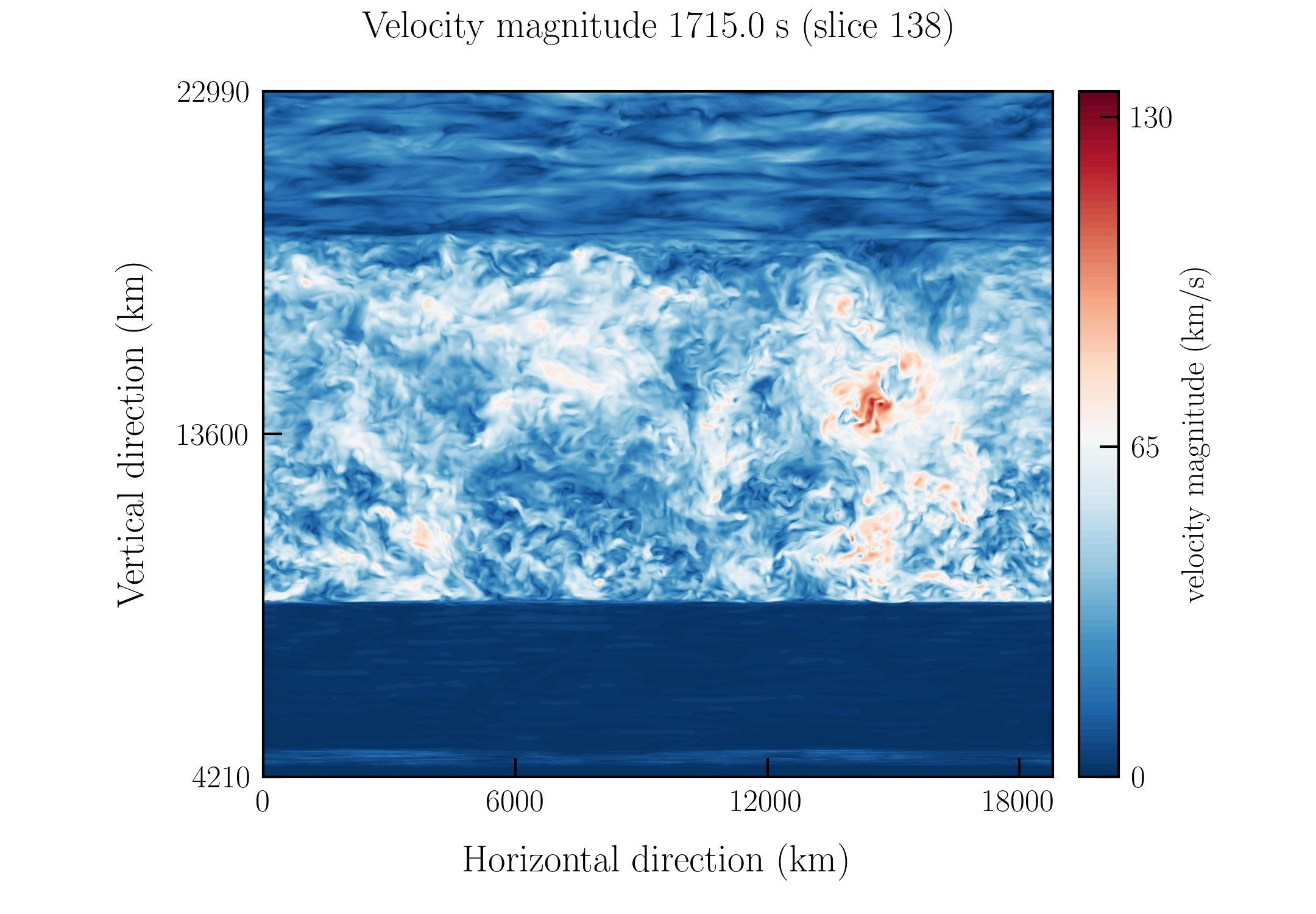}
\caption{Vertical slice of a 3D, $1024^3$ simulation  of a carbon burning shell \citep{andrea}. Velocity magnitudes are shown (red is high; white is medium; blue is low). This figure illustrates the complexity of the turbulent flow, and strikingly shows the boundary layers which form at top and bottom. The structures---rolls and plumes---are intermittent: forming, breaking apart, and re-forming elsewhere. Both boundary layers are also dynamic, and radiate g-mode waves (most clearly visible at the top).
}
\label{fig1}
\end{figure*}
\placefigure{1}

Our set of 3D numerical simulations\footnote{See  \cite{ma07b,amy09vel,am11turb,viallet2011,321D,andrea,andrea2,miro}.}
are ``box-in-star" computations which range from ``very low resolution'' ($128^3$ zones) to  ``very high resolution"  (now $1536\times 1028^2$ and $1536^3$ zones). These are implicit large eddy simulations \citep[][ILES]{woodward,iles07,apsden}. They include two stable layers sandwiching a turbulent convective region; see Fig~\ref{fig1}, which shows a cross-sectional slice through a representative case. 
 Velocity magnitudes are shown (red is high; white is moderate; blue is low). This figure illustrates the complexity of the turbulent flow, and strikingly shows the boundary layers which form at top and bottom of the turbulence. The boundary layers are thin, and not step functions but sigmoids \citep{andrea}.
 Coherent structures---rolls and plumes---are strongly dynamic: forming, breaking apart, and re-forming elsewhere. Both boundary layers are also dynamic. They bend and stretch, and radiate g-mode waves (which are most clearly visible at the top).
 There is intermittency \citep{tennekes} in both space and time, as the ``patchiness'' in Fig.~\ref{fig1} indicates;  intermittency is related to nonlinear interactions of coherent structures \citep{warhaft}.
 
 This is confirmed by movies of (1) the evolution in time (``Very high resolution movie of the C-shell''), and of (2) a fly-through of the computations at a given instant in time (``Carbon shell ($1024^3$) simulation: fly-through movie").\footnote{The time evolution and the fly-through  movies may be found at  
 \url{http://www.astro.keele.ac.uk/shyne/321D/convection-and-convective-boundary-mixing/visualisations}.} 
To the extent that the simulations are in a statistical steady state in time, and statistically homogeneous in space, the movies will have a similar visual appearance (as they do). Turbulence has a space-time isotropy. This allows averaging procedures to be robust.

Unless stated to the contrary, our simulations include a full reaction network of appropriate size for the problem at hand \citep{wda96}, with reaction rates coupled to the thermodynamic and compositional fluctuations, so that nuclear heating and neutrino cooling are dynamic variables. Simplifications are only made after explicit testing of their validity. 

\subsection{Methods: ILES and RA}\label{Smethods}

The simulations extend in resolution from above the integral scale of turbulence, down to well inside the inertial range of the cascade \citep{andrea}. 
The full cascade is represented because the method used (PPM, \cite{cw84})  solves the Riemann problem for nonlinear flow at the individual zone level. This method and others of its class (finite-volume monotonic solvers, \cite{iles07,leveque}) automatically result in a Kolmogorov description of the turbulent cascade down to a dissipation scale at a sub-grid level.
These ILES methods  use Riemann solvers to capture shocks at the scale of a zone $\Delta r$. They relate  the change in turbulent kinetic energy $(\Delta v)^2$ and traversal time $\Delta r / \Delta v$ across the shock,  to the rate of specific entropy production across the shock $-{ \partial S \over \partial t} \sim (\Delta v)^3 /\Delta r$. This is the Kolmogorov expression for a turbulent cascade, so these methods automatically (implicitly) match motion to a turbulent cascade at the grid scale.\footnote{See Ch.~2 in \cite{iles07}.} 
 Conservation of mass, momentum and energy are enforced to machine accuracy, so that numerical error is concentrated in the calculated shape of the flow field, not the conservation laws. 
    
 These simulations solve the Euler equations, not the Navier-Stokes equations,  so the question of convective instability is determined by the Reynolds number. The formal Rayleigh number may be infinity (no explicit radiative diffusion\footnote{Some simulations had realistic radiative diffusion added, but it was too small to have a noticeable effect.}). The 1D model used for guidance
\citep{andrea} had a very high P\'eclet number (Pe $\sim 10^8$), but the turbulent cascade implies that the results are insensitive to  Pe (see also \cite{orvendahl}). The convective Mach numbers are small ($< 0.02$).

For these simulations the effective Reynolds numbers are  roughly $Re\sim n^{4/3}$, where $n$ is the  number of zones across the turbulent layer, so 
$Re \sim 600 $ to more than $2 \times 10^4$. This allows us to compute from the integral scale of fully turbulent flow, down into the inertial range of the cascade\footnote{We explicitly confirm that this range is attained, using both  turbulent velocity spectra and dissipation rates; see below.}, with no additional assumptions concerning flow geometry.

To tame the turbulent fluctuations we integrate over angle (or the horizontal dimension of the convective region), and over several turnover times, to obtain the average behavior over a spherical shell. 
A novel and key feature of our procedure is that it avoids the classical closure problem of the Reynolds-averaged Navier-Stokes (RANS) equations \citep{tritton}.
The numerical simulations do not have the un-physical non-dynamic fluctuations found in unconstrained statistical methods, and therefore provide exact averages, limited only by the granularity of the space-time grid. To be precise we call our method ``Reynolds-Averaged ILES'', or RA-ILES \citep{miro}. 
The RA-ILES method allows an accurate and separate assessment of dissipation due to (1) turbulence, and (2) resolution error, as we show in the companion paper.
 Comparison of simulations with different resolution shows the effects of finite zoning.
 For more detail 
 see   \cite{321D,viallet2011,andrea,andrea2,miro}.

 The  ILES approach is a natural complement to  the more traditional method, Direct Numerical Simulation (DNS), which resolves the dissipation scale \citep{pope} but strains to encompass the larger scales, and generally cannot deal with turbulence at those scales which generate motion in stars. In contrast, ILES resolves these large scales, but approximates by a  Kolmogorov cascade the behavior downward  to a much smaller dissipation scale. Kolmogorov theory is one of the success stories of turbulence 
 \citep{frisch}, but it too is not perfect  as it does  not contain the direct influence of the large scales on small scales \citep{warhaft}. In ILES this is accomplished at least in part by the nonlinear interaction of large scales, giving intermittency (\cite{tennekes,holmes}, \S\ref{Ssims}).
DNS and ILES have different strengths and weaknesses, so that taken together they provide  a fuller picture. 

    RA-ILES adds to ILES the power to detect errors due to finite granularity in the computational  spacetime, and allows us to quantify the relative importance of different physical effects in our problem \citep{miro}. This aids in the construction of mathematical models of much lower complexity, whose solutions nevertheless approximate the full numerical simulations. We shall  freely use such approximations to illustrate the underlying physics. Thus, our procedure has three components: numerical (ILES), analysis of numerical simulation (RA), and analytic approximation (321D).

\subsubsection{Limitations: rotation and MHD}
Even a perfect solution to  the B\"ohm-Vitense problem would not solve the more general issue of convection in stars.

\cite{featherstone} suggest that supergranulation is a rotationally-constrained flow; to add rotation goes beyond the B\"ohm-Vitense formulation.
Rotation requires a  star-in-box approach to capture the largest scale; see
pioneering simulations of \cite{porterwoodward}. 
Rotation forces non-locality \citep{am10rot} and symmetry breaking \citep{viallet2013}.

 To the extent that magnetohydrodynamics (MHD) is important, the symmetry is broken upon which Kolmogorov (and we) rely; that is an issue for future research\footnote{A preliminary step \citep{am10rot} suggests that mild rotation causes turbulent flow to tend toward conservation of specific angular momentum, with dissipation provided by the turbulent cascade.}.
Turbulence makes and stretches vortices (e.g., \cite{pope}); a seed magnetic field will be compressed (doing work against magnetic pressure) and stretched (doing work against magnetic tension);
\cite[see][]{parker,davidsonmhd}. Fluid kinetic energy is therefore converted into magnetic field energy, and fluid flow is retarded, giving a dynamo. 
Magnetic fields are buoyant and will tend to rise in a gravitational field. Unless all the field escapes, it strengthens.
Stronger magnetic fields tend to stop the flow, which then no longer generates magnetic field, allowing convection to reassert itself. There is a tendency for a magnetic cycle, reminesent of the solar cycle.

Stars are made of high energy-density (HED) plasma\footnote{See \cite{dqlB} for a recent experimental attempt on the Omega laser to address this problem.}, so that magnetic fields will be ubiquitous in stars, but what geometry, strength  and dynamic behavior will they have?
Geometry is evidently important: 2D turbulent flow develops a reverse cascade in which strong vortices form, merge, and grow in size, while in 3D such vortices are shredded.

We have chosen the simplest version of this complex problem, that of
non-rotating, non-magnetic B\"ohm-Vitense convection.
We have added turbulence, stratification, non-uniform composition, time dependence, non-locality, and boundary physics, but
incorporating rotation and magnetic fields remains a challenge for the future.
We regard our RA-ILES method as an interesting approximation which, unlike DNS,  can resolve the integral scale for stellar turbulence.

\section{Some insights from the Turbulent Kinetic Energy Equation}

Simulations of 3D turbulence have a very large number of degrees of freedom; to understand them we will frequently use mathematical models which have solutions similar to the 3D simulations, but  a much reduced set of degrees of freedom. In spirit this reduction parallels the method of \cite{holmes}, but in contrast uses RA-ILES analysis to identify dominant terms (\S\ref{Smethods}). To better understand these results we construct even simpler analytic solutions, more at the level of  (but better than) MLT.

\subsection{Blobs?}
One major flaw in MLT is the 
idea of a convective ``blob'', which is created,  accelerated by buoyancy along a ballistic trajectory of   length $\ell$, and then dissolved into the background. As \cite{alvio87} discusses, this causes problems at the  beginning and  the end of the trajectory (the {\em ends} problem), and introduces a free parameter $\ell$, which is taken to be {\em both} the size of the blob and the distance it travels.

No such ``blob"  behavior is seen in 3D simulations\footnote{Rayleigh-Taylor blobs may appear briefly at the start of convection or explosion.} of stellar atmospheres \citep{sn89}, or interiors \citep{porterwoodward,ma07b}; instead the behavior is that of a complex and dynamic combination of rolls, plumes and other shapes (Fig.~\ref{fig1}).  Renzini's problems 1, 2, 3 and 4 in \S\ref{s-alvio} are all related to flaws in this ``blob"  concept.

\subsection{Length  scale for convective velocity}\label{S-mass}

Baryon conservation and the assumption of a quasi-static, quasi-spherical background place strong general constraints on the nature of convective flow in stars:  mass flux balance
\begin{equation} 
\bf \nabla \cdot {\rm \rho} v = 0 \label{eq-mass-flux}
\end{equation}
implies, for a spherically symmetric star,
\begin{equation}
{\bf \nabla \cdot v} =  v_r/H_{\rho}, \label{eq-dens-strat}
\end{equation}
where $H_{\rho}$ is the density scale height\footnote{A related length, the {\em pressure} scale height is usually used in MLT, 
where $H_P = -(
 \partial \ln P /\partial r )^{-1}$ .} defined as $-(\partial \ln \rho / \partial r)^{-1}$. 
Eq.~\ref{eq-dens-strat} connects the convective velocity structure to a length scale without any MLT assumption.
For a medium of uniform density, $H_{\rho} \rightarrow \infty,$ and the scale becomes the size  of the turbulent medium, which is finite. 
This does give the characteristic {\em length scale} for a quasi-steady flow in a stratified medium (such as stellar convection). 
However, this equation is linear in velocity, so that the  {\em velocity scale} is not constrained; another equation is required to determine it 
(involving convective enthalpy flux, for example).

Any form of ``rotational'' flow (Eq.~\ref{eq-mass-flux})  can remove the {\em ends} problem 
because such flow turns back to remain in a finite volume. 
\cite{lorenz} showed that the simplest version of such  a flow (a 2D convective roll) is an example of deterministic chaos; the flow chaotically flips from clockwise to counter-clockwise. In 3D the angular momentum vector is not restricted to only two orientations (as in 2D), but may wander through $4\pi$ steradians\footnote{See \cite{am11turb} for a roll model, and also \cite{gabriel18} for a plume model.}. Such instabilities associated with one or several strange attractors seem to act as seeds for turbulence (see \S5). 

\subsection{Length scale for turbulent damping}
The viscosity $\nu$ of stellar plasma is roughly comparable to that of common fluids, 
but stars are so large that stellar Reynolds numbers\footnote{The Reynolds number is the product of the length and velocity scales, divided by the viscosity \cite[][Eq. 19.1]{llfm}.} are far larger than we encounter terrestrially, making their flows highly turbulent \citep{am16key}. 
\cite{kolmg,kolmg41} showed that  $\epsilon_K$, the rate of flow of turbulent specific kinetic energy through a 3D turbulent cascade, is determined by the large scale flows (\cite{llfm}, \S31), 
\begin{equation}
\epsilon_K \approx \overline{ v^3}/\ell_d, \label{kolmogorov}
\end{equation}
where $v$ is the speed, $\ell_d = \alpha H_P$ the linear size of such flows, and the transit time is 
$\tau = \ell_d/|v|$. 
The flows at such  scales are affected by boundary conditions, so that $\alpha$ is not a universal constant, and Eq.~\ref{kolmogorov} may apply only on average \citep{amy09vel}.
This corresponds to a drag (negative acceleration) of
 \begin{equation}
{\cal D} = -{\bf  v} |v| / \ell_d  = -{\bf v}/\tau \label{eq-drag}
\end{equation}
 \citep{321D}. The mixing-length assumption provides a representation of the effect of a drag term, but no estimate of its value.

\subsection{Balance between driving and damping}\label{balance}

The chaotic driving of turbulence causes large fluctuations, and requires that we average instantaneous properties to obtain useful variables for stellar evolution (\S2.6 in \cite{321D}, and \cite{ma07b}). Fluctuations are not part of MLT; see the {\em fluctuations} problem in \S\ref{s-alvio}.
We do a double average, over angles (spherical shells), and over several turnover times, 
\begin{equation}
\tau_{to}=2  \Delta r_{cz} /v  = 2\tau , \label{eq-turnover} 
\end{equation} 
which are short compared to evolutionary times\footnote{For hydrogen and helium burning the turnover time $\tau_{to}$ is so much shorter than the nuclear burning times that the convective algorithm may be ``stiff''\citep{acton}, and require special attention.}. Here $\Delta r_{cz}$ is the depth of the convection zone.

When performed on  even modestly resolved numerical simulations of  
convection, 
such  averaging shows a balance over the turbulent region, between  (1) large scale driving 
and  (2) dissipation at the small-scale end of the turbulent cascade. 
Convection in strong stratification ($\Delta r_{cz} \gg 2 H_P$) is also driven by ``pressure dilatation'' as well as buoyancy \citep{viallet2013}. 

For weakly-stratified convection zones (and MLT) the buoyancy driving (the work done by buoyant acceleration  $\bf {\cal B}$ acting on the turbulent velocity
$\bf v$) is 
 \begin{equation}\label{eq-vbB}
{\bf v \cdot }{\cal B} = {\bf v \cdot g} \beta_T \Delta \nabla ,
\end{equation}
where  the gravitational acceleration vector is $\bf g$, the super-adiabatic excess is
$\Delta \nabla= \nabla - \nabla_e ,$
and $\beta_T $ is  the thermal expansion coefficient \citep{kippen}. 
The entropy excess $\Delta \nabla$ may contain  contributions from composition differences, which  are ignored in MLT. To evaluate these, the issue of mixing must be solved consistently with that of convection, complicating the problem \citep{321D,woodward2015,miro}.

The rate of dissipation for turbulent kinetic energy due to the turbulent cascade is, on average,
 \begin{equation}
\overline{{\bf v \cdot }{\cal D}} \approx \epsilon_K, \label{eq-kol}
\end{equation}
which is essentially the Kolmogorov value for homogeneous isotropic turbulence\footnote{For our purposes this may be adequately accurate, but the idea of homogeneous isotropic turbulence may be over-simplified; intermittency and anisotropy observed  at small scales may require refinement of these ideas \citep{warhaft}. See \S\ref{Smethods}.} \citep{frisch}.

The Brunt-V\"ais\"al\"a frequency squared  \cite[Eq.~6.17,][]{kippen} is
\begin{equation}
{\bf{\cal N}^2} = -{\bf g} \beta_T \Delta \nabla /H_P,
\end{equation}
so that we may write an equation
\begin{equation}
d{\bf v}/dt = \partial {\bf v}/\partial t +{\bf v \cdot \nabla v} =  -H_P {\bf {\cal N}^2} - {\bf v}/\tau \label{mlt_acc},
\end{equation}
whose solutions can be related to the simulation results.
For $-H_P {\bf {\cal N}^2} >0$ we have convective flow which is turbulent.
This is basically a statement of Newtonian mechanics, with driving by buoyancy and damping by drag\footnote{If ${\cal N}^2 \gg 0$ we are in the wave regime, so that ${\bf v}/\tau$ must be negligible, and we have the wave equation. For 
${\cal N}^2 \approx 0$ but not 
${\cal N}^2 \gg 0$, 
buoyancy braking may be operative, and Kolmogorov damping happens. This is a surface of separation, where Prandtl theory has a singularity.}. 
\cite{gough77} gives a historical context going back to Prandtl and to Biermann. The early attempts (and most of the recent ones) have used a kinetic theory model, in which the mixing length was a sort of mean free path. To connect with numerical results, we prefer a model of the momentum equation for fluid dynamics, involving structures such as waves, convective rolls, or plumes.

Taking the dot product of Eq.~\ref{mlt_acc} with $\bf v$ gives a turbulent kinetic energy equation,
\begin{equation}
d(v^2/2)/dt = {\bf v \cdot g} \beta_T \Delta \nabla - v^2/\tau, \label{mlt_KE}
\end{equation}
for which the steady-state solution\footnote{Care must be taken (for negative $v$) with the sign of the transit time $\tau$ and the deceleration.} is a balance between driving and damping, with
$\ell_d \equiv  {\ell^2_{MLT} / 8 H_P }$ for MLT,  and $\Delta \nabla >0$. 
 In Eq.~\ref{mlt_KE}, negative values of $\Delta \nabla$ are allowed; this permits buoyant deceleration. The singularities in MLT at the convective zone boundaries (\S9 in \citealt{gough77}), and in boundary layers (\S40 in \citealt{llfm})    are removed by the introduction of the total time derivative of the specific turbulent kinetic energy. 
 The singularities in this case occur in Prandtl's equations for a boundary layer as the velocity perpendicular to the surface goes to zero. In a star the motion does not go to zero but becomes wave-like\footnote{Because the Mach numbers are low, gravity (g-)modes dominate over compressional (p-)modes.} rather than turbulent (e.g., \S\ref{Swavegen}; Fig.~\ref{figo16}).

The flow is relative to the grid of the background stellar evolution model, so 
the co-moving time derivative of turbulent kinetic energy may also be written as
\begin{equation}
d(v^2/2)/dt = \partial_t ({\bf v \cdot v})/2 + {\bf \nabla \cdot F_{K} },\label{FKE}
\end{equation}
where $ {\bf F_{K} } = \rho {\bf v} ({\bf v \cdot v})/2$ is a flux of turbulent kinetic energy. The generation of a divergence of the kinetic energy flux in this way is robust for dynamic models; it occurred in the precise RANS approach as well \citep{ma07b}. 
This flux acts to spread locally-driven turbulence more evenly over the turbulent region, to be dissipated as Kolmogorov suggested. Eq.~\ref{mlt_KE} and \ref{FKE} together comprise a spartan form of the turbulent kinetic energy equation \citep{ma07b}.

Our RA-ILES numerical simulations  satisfy Eq.~\ref{eq-kol} and \ref{mlt_acc}
\citep{amy09vel,am11turb} 
 and as well as a local balance on average, of driving by acceleration and deceleration,
\begin{equation}
\overline{ {\cal B}} \approx \overline{{\cal D}}, \label{eq-balance}
 \end{equation}
over a weakly-stratified convection zone  \citep{ma07b,321D}; phase lags between driving and damping cause the pulses seen in Fig.~\ref{fig-evol} (below). 
 
 Thus, on average,
\begin{equation} 
 v^2/\ell_d \approx g \beta_T \Delta \nabla , \label{eq-v2} 
 \end{equation}
which is something like MLT: see  \cite{321D} 
for a discussion which allows buoyancy braking. This equation is nonlinear in velocity and therefore can set a velocity scale for Eq.~\ref{eq-dens-strat},
but {\em with the Kolomogorov damping length $\ell_d$  replacing 
the mixing length.} 

Use of the turbulent cascade removes the {\em two lengths} problem. 
{\em
The damping length is not a size, but a measure of the rate of damping due to turbulence. }
 Further, the cascade resolves all scales down to the  dissipation scale of Kolmogorov, so that the {\em resolution  problem} also disappears.
 
 These results are reminicent of some previous work, e.g.,
\cite{cm91,can96} who developed a theory of full-spectrum turbulence, and
\cite{can11a,can11b,can11c,can11d,can11e} who made further progress with a
Reynolds-stress approach. These earlier suggestions, like the 3D simulations, shift the focus from the original MLT picture 
of blobs to one involving a turbulent cascade. 

\section{Evolution of turbulent kinetic energy}\label{S-evolTKE}

\begin{figure*}[h]
\figurenum{2}
\label{fig-evol}
\includegraphics[angle=0,scale=0.23]{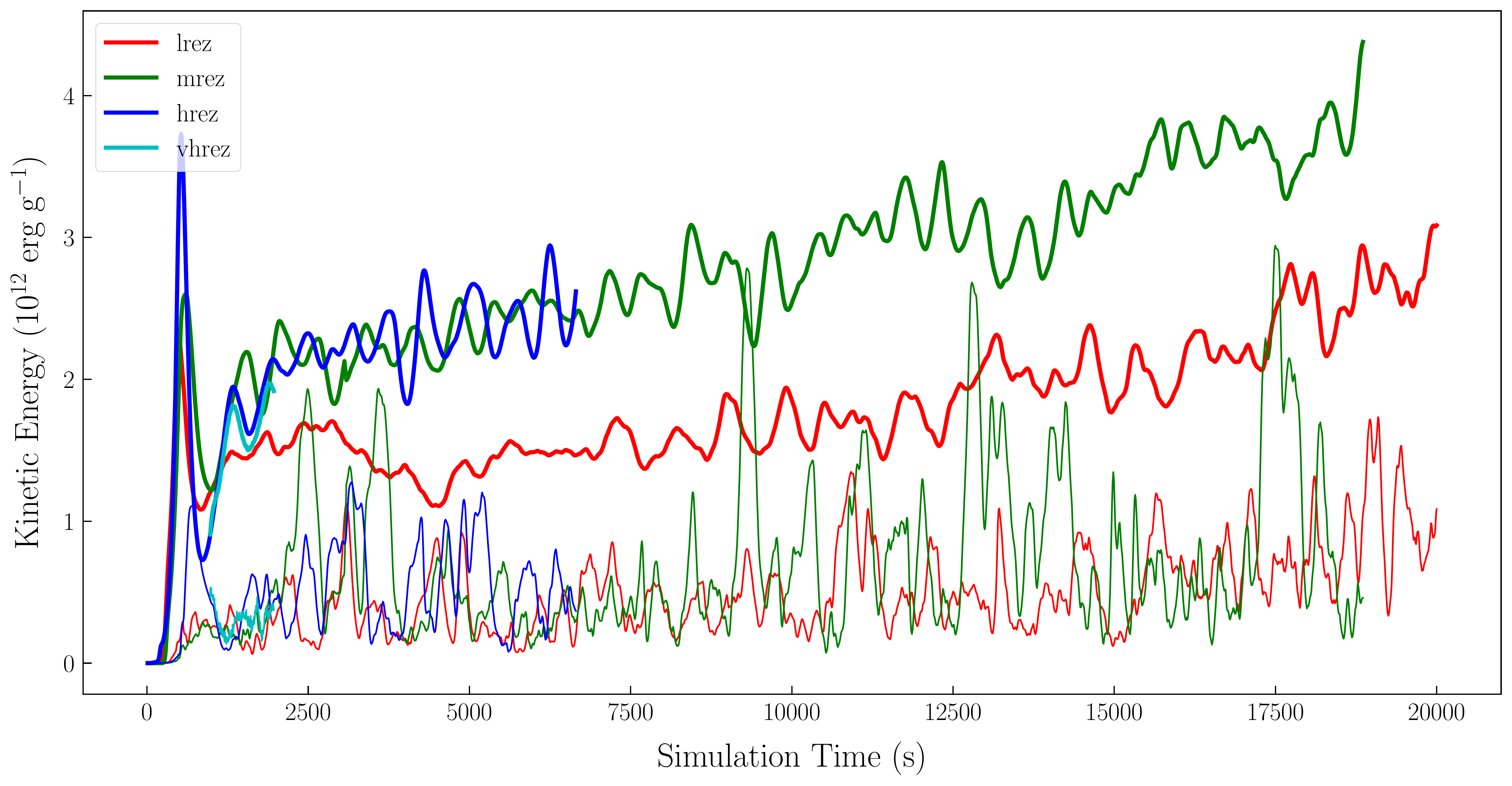}

\includegraphics[width=1.1\textwidth,height=0.7\textwidth,trim=120 0 0 0,scale=0.6]{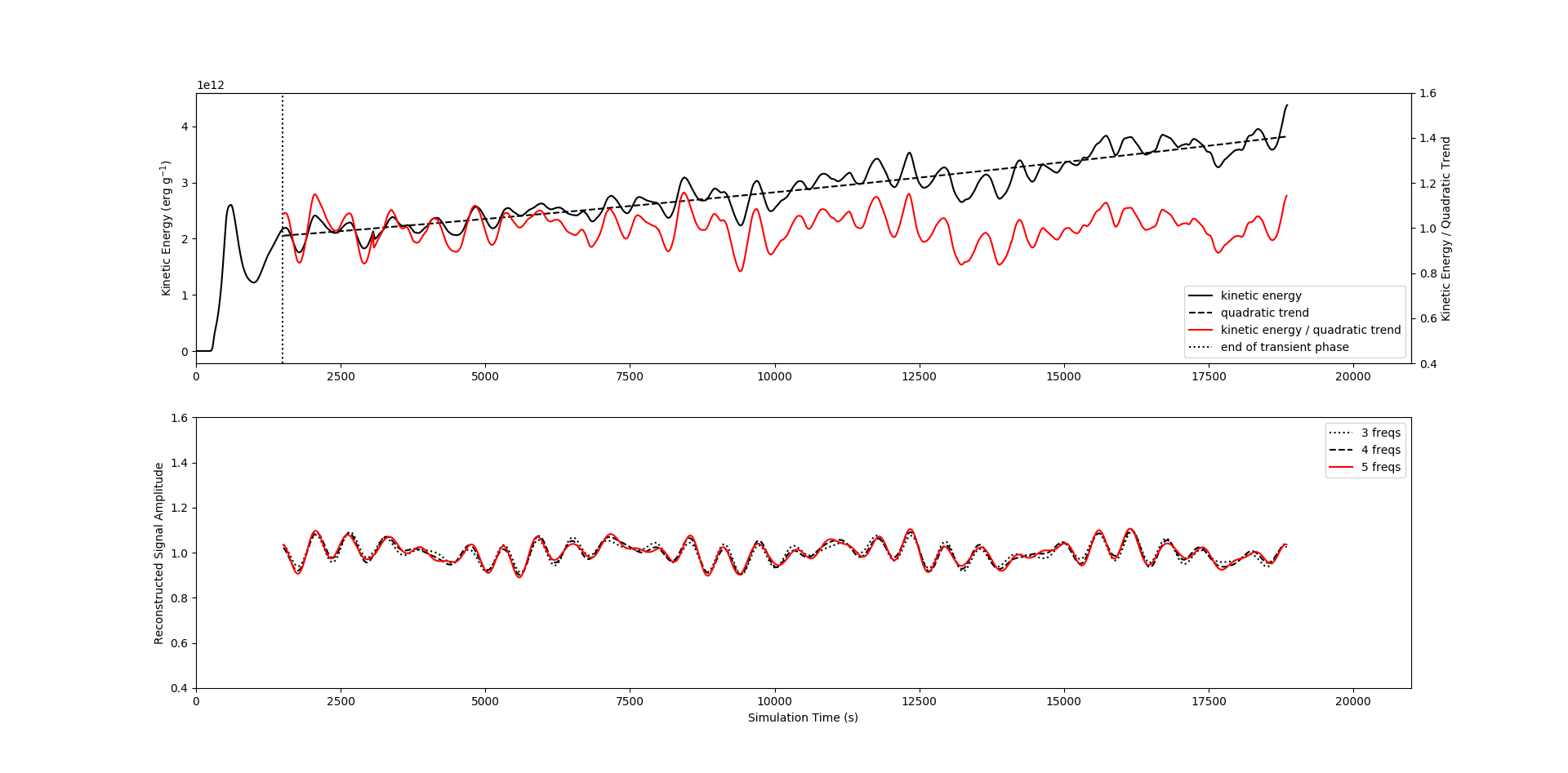}

\caption{({\it top pane}) Specific turbulent kinetic energy (TKE)
for the carbon burning shell \citep{andrea,andrea2}, versus radius for 18,000 seconds. This shows ``lrez'' ($128^3$, red 
dots) and ``mrez'' ($256^3$, green 
dashes),  and  "hrez" ($512^3$, solid blue 
line) simulations. 
Fluctuations in TKE are seen, as found by \cite{ma07b}. Multi-mode behavior is evident.
The $128^3$ simulation shows effects of lower resolution (higher numerical viscosity), but $256^3$ and $512^3$ seem unaffected. Wave kinetic energy (same colors but thin lines) is much smaller, and is plotted (multiplied by a factor of $25$) along the bottom. Higher resolution simulations ($768^3$, $1024^3$, and $1536^3$ have not been done for such long times, but agree fairly well with the $256^3$ and $512^3$ cases shown. A small segment in cyan of the $1024^3$ simulation is shown for illustration.
({\it bottom two panes}) Mode analysis of $256^3$ simulation, for frequencies  1.6, 0.23, 1.8, 0.81, and 2.1 mHz. A trend of $(5.26 \times 10^3 t^2  -­ 4.41\times 10^7 t + 1.56\times 10^{12} )$, $t$ in $s$, was divided out; it corresponds to an evolutionary change (see text). 
While the pulses are not sine waves, they are periodic, so we separate out a few (5) frequencies that are shorter than the turnover time, as expected for a turbulent cascade. Even five modes begin to capture the "pulses" moderately well (bottom panel). These low order, multi-mode fluctuations are robust features of the TKE.
}
\end{figure*}
\placefigure{2}

Kolmogorov dissipation is derived for a homogeneous, isotropic turbulent medium \citep{frisch},
where boundaries are not important. We accept this hint, and first examine the behavior of specific turbulent kinetic energy in bulk, reserving a discussion of boundaries until later (\S\ref{s-bnd}).
 
How much kinetic energy is involved in the convective motion? This is Renzini's {\em flux of turbulent kinetic energy} problem.
Fig.~\ref{fig-evol}, top pane, shows the evolution in time of specific turbulent kinetic energy (TKE) in the carbon-burning simulations \citep{andrea,andrea2}, for resolutions of $128^3$, $256^3$, and $512^3$. Multi-mode behavior, like that  seen in \cite{ma07b} (Fig~4) and \cite{amy09vel} (Fig.~5), is evident. 
Simulations of $768^3$, $1024^3$, and $1536^3$ were not continued for such long times, but they tracked the $256^3$ and $512^3$  results; the short $1024^3$ case is shown in cyan.
All simulations are consistent with an approach to a quasi-steady state, but with {\em significant and continuing fluctuations} around that average value. Such behavior, while typical of turbulent flows, is not in MLT (Renzini's {\em fluctuation problem}), but is a feature of high resolution simulations; see also Fig.~5 in \cite{woodward2015}.

\subsection{Initial transient}\label{SinitKE}
Each simulation is initiated from a hydrostatic state, recalculated on the grid to machine accuracy. It is then overlaid with very small  random perturbations in density, with amplitudes which were $10^{-3}$ of the initial static value. Convection grows gently in the unstable region, gradually forming a non-linear chaotic flow (the turbulent cascade plus coherent structures). The convective Mach numbers rise to $\sim10^{-2}$; these kinetic energies are $10^4$ times larger than implied by the initial seeds. Such small perturbations quickly disappear in the stable regions\footnote{For the C+C shell simulations, the perturbations were only put in the convective layer. In the O+O simulations the perturbations were put in stable layers too but quickly disappear from these regions.}. 

Initially there is no turbulent dissipation, only driving by buoyancy.
A large first pulse develops due to this delay in the turbulent cascade\footnote{Because the first pulse also depends on interactions between multiple modes of flow, there is no guarantee of simple behavior.
However, compare Fig.~\ref{fig-evol} to the very similar Fig.~5 in \cite{woodward2015}, which represents apparently quite different physics: H entrainment at the top of He-shell flash.
}. After this pulse, driving balances dissipation on average, but not exactly:  there is a phase lag  \citep{am11turb}, so that {\em the pulses do not disappear because damping always lags driving.}\footnote{This computational domain was chosen to study separate convective cells; an average over a larger domain (e.g., $4\pi$ steradians) would contain more chaotic cells, and not resolve the pulse behavior so well.}

There is an evolutionary growth in TKE due to a slight mismatch of thermal balance between initial conditions based on a 1D MLT model, and the energetically scaled 3D model \citep{andrea}. The heating was $10^3$ of its realistic value; although speeding the evolution correspondingly, the heating per turnover is still very small in comparison to the internal energy. This trend is shown as a line in the middle pane, and is removed in order to extract frequencies of the dominant modes (bottom panes).  A few (five) modes are sufficient to capture most of the multi-mode behavior, as suggested by \cite{holmes}.

\subsection{Multi-modal behavior}\label{Smultimode}
Movies of the simulations support the view that the fluctuations in Fig.~\ref{fig-evol} 
are caused by turbulent break-up of multiple 3D rolls. Such multi-modal behavior also causes motion of the convective boundaries, driving waves into neighboring regions, and causing variations in TKE. 
These pulses are clearly seen  in Fig.~\ref{fig-evol}, and do not attenuate noticeably over $\sim 30$ turnover times.

Oxygen burning requires no scaling of the heating rate because the consumption of fuel may be explicitly followed. However, the interaction of turbulence, burning, and mixing of Ne and O is more complex than generally realized \citep{miro}, so that we simplify at this point by focusing on  C burning, which is similar but has no Ne ingestion. On these time scales ($2 \times 10^4\,$s), little carbon is consumed, even with the enhanced rate. These carbon burning simulations need no explicit nuclear evolution over this time scale.  

The middle pane in Fig.~\ref{fig-evol} shows the original TKE curve and a flatter one after ``detrending'' to remove the effect of a slow thermal evolution (see caption).
The bottom pane shows the reconstruction for 3, 4, and 5 frequencies, which resembles the ``detrended'' curve increasingly well; at ten frequencies (not shown) the fit is excellent, but even a few modes are sufficient to capture the basic behavior.
A proper orthogonal decomposition should allow a better representation \citep{holmes}, but for now we simply want to emphasize the robust nature of the pulses in time.
The pulses, which are comparable to a transit time in duration, imply that statistical estimates of turbulent properties 
 will be modulated by multi-mode behavior, unless averaged over many transit times.

 \subsection{Wave generation}\label{Swavegen}
The kinetic energy in waves is shown, in the top pane in Fig.~\ref{fig-evol},  as thin lines, scaled up by a factor of 25 for better visibility and using the same color code as for resolutions. 
\cite{321D} showed that the boundary of  convection has a particular, dynamically required structure, which implies a particular rate of wave generation.
In order for matter to turn and remain in the turbulent zone, outgoing flows must be decelerated. Buoyancy braking requires that the  square of the Brunt-V\"ais\"al\"a frequency  change sign (Eq.~\ref{mlt_acc}). This means waves are 
supported, with their amplitude depending upon the stellar structure and the vigor of convection. The buoyancy braking provides a direct connection between convective and wave motion.
Thus, the physics of a convective boundary {\em requires} the generation of waves.
In this case the energy in the waves is small relative to that in convective motion.
See \S\ref{s-bnd}, Fig.~\ref{figo16}, where at $r < 0.43 \times 10^9$\,cm, the blue curve gives evidence for a background of g-mode waves.

\subsection{Linear Stability Theory}
Fig.~\ref{fig-evol} contains another implication for stellar physics.
In {\em linear} stability theory  \citep{aerts,unno} it is not possible to include a realistic
treatment of convective driving and damping because these terms are  inherently 
{\em non-local} and {\em non-linear} (see discussion of the $\tau$-mechanism in \cite{am11turb}). This issue is sometimes called the ``time-dependent convection'' problem.
Actually,  {\em all} stellar convection is ``time-dependent'', to  the extent that it based on turbulent flow. Because of intermittency, convection  is only ``steady-state'' in an average sense, if at all.
The  multi-modal behavior of the simulations is most simply described as an interaction of a few chaotic 3D convective rolls (\cite{lorenz,holmes,am11turb}, and Fig.~\ref{fig-evol}).
Turbulent convection drives waves which, once launched, may then be described by linear theory.
A sequence of increasingly stronger waves (increasing convective Mach number) transforms pulsations into explosions.

\subsection{Resolution}
With each new initial resolution, synchronization is lost, so that each simulation is an independent member of a ensemble; all of which are attracted to the cascade in the long term. The RA-ILES
integrated values are reproduced with surprising accuracy even at crude resolution ($128^3$), and are well resolved at ($256^3$) and above \citep{ma07b,viallet2013,321D,andrea,andrea2}. 

The lowest resolution case ($128^3$) has the highest numerical dissipation; it settles toward a quasi-steady state with the lowest TKE in Fig.~\ref{fig-evol}. This higher dissipation may also be responsible for the reduced amplitude of the TKE peaks in $128^3$   relative to, e.g., the $256^3$ simulation. 
While high resolution and long evolution are both desirable, they are in conflict for a finite computer budget, so the highest resolution runs are relatively short.

\section{Composition and boundaries}\label{s-bnd}

The study of turbulent kinetic energy has led us back to the issue of convective boundaries, which is not a part of MLT, but which we now address.

\subsection{Errors in dynamics at stellar boundaries}
Boundaries of stellar convective regions are assumed to be adjacent to regions of convective stability (``radiative" regions assumed to have zero flow velocity). {\em Thus convective flow must be joined to zero flow,} which leads to the infinite accelerations and decelerations in MLT (the {\em ends} problem of \S2). 

The treatment of boundaries requires a consideration of dynamics not included in MLT. Because of fluctuations, convective flow at a boundary can only be zero on average. This implies that the boundary moves, and is a source of waves. Pure radial motion gives compression, and p-mode waves. However, non-radial motion is also nonzero, giving shear and gravity-modes.
At low Mach numbers the g-modes dominate; only these are visible in Fig.~\ref{fig1}.

Terrestrial experiments and numerical simulations show that convective and non-convective regions are separated by a boundary layer (Fig.~\ref{fig1}), which is required to join rotational (${\bf \nabla \times v } \neq 0$) and potential (${\bf \nabla \times v } = 0$) flow patterns (\cite{llfm}, \S44). 
Rotational flow is associated with mixing and potential flow is not. Such layers require large gradients to perform the joining of these very different flows, and thus the layers are thin.  
These layers provide buoyancy braking to turn the flow, which implies negative buoyancy and thus a Brunt-V\"ais\"al\"a
frequency which is real (\cite{lighthill}, \S4.1), so that these layers support waves (see Eq.~\ref{mlt_acc}). They are well mixed by the turbulent flow.
In such layers convective fluctuations must be coupled to wave production. 
Beyond these layers,  in the radiative zone which may support a composition gradient,
the Brunt-V\"ais\"al\"a frequency is also real, so internal waves are supported,
and directly coupled to the boundary layer fluctuations. Because they can carry vorticity, g-mode waves can induce some energetically-limited mixing beyond the boundary layer.

{\em ``One of the properties of the region of rotational turbulent flow is that the exchange of fluid between this region and the surrounding space can occur in only one direction. The fluid  can enter this region from the region of potential flow, but can never leave it."} \citep[][\S34]{llfm}. 
 Mixing, which increases entropy, gives the uni-directional nature of the exchange. 
Such {\em entrainment of material from a radiative zone} is a necessary aspect of convective boundaries,  and one not included  in MLT. 
This does not preclude receding convective regions, which could shed regions of decaying turbulence; 
also see \cite{holmes}.

\subsection{Distinct boundary regions}

\begin{figure*}[h]
\figurenum{3}
\label{fig-toon}
\includegraphics[trim=-10 540 -80 0,scale=1.4]{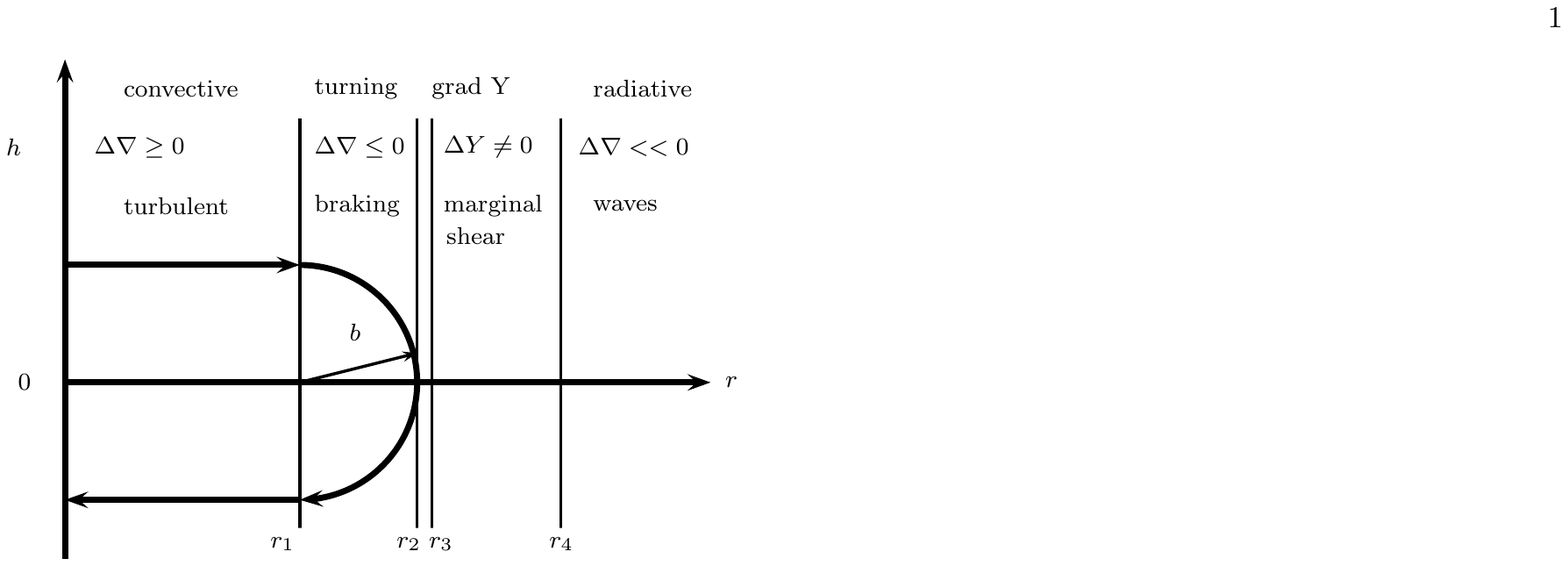} 
\caption{Two-dimensional schematic of  the average structure of a convective boundary. The length $b$ corresponds to the radius of curvature needed to reverse (contain) the flow ($v_r \rightarrow - v_r$). The centrifugal acceleration is provided by negative buoyancy.
The radial direction is denoted by $r$ and the transverse by $h$. 
The boundary layer lies between $r_2$ and $r_3$. Imagine the whole system undulating due to turbulent fluctuations.
}

\end{figure*}
\placefigure{3}

We construct a cartoon to clarify discussion of the average properties; see Fig.~\ref{fig-toon} which shows the boundary of a convection region (convective to the left, radiative to the right).
There are five distinct regions, only two of which (1 and 5) are recognized in MLT:
\begin{enumerate}

\item{Convection.} Inside $r_1$, the superadiabatic excess is positive, as are the buoyant acceleration and the enthalpy flux. In this region there is a balance between buoyant driving and turbulent dissipation  (\S\ref{balance}). This is the ``convective'' region of MLT \citep{bv58}.

\item{Braking.} Between $r_1 $ and $r_2$ the superadiabatic excess becomes negative, and the flow can turn, and reverse direction. This is the ``braking'' region which is required dynamically. It has a negative enthalpy flux, is well mixed, and generates waves. This region does not exist in MLT, and is discussed in \cite{321D}. 

\item{Shear.} Between $r_2$ and $r_3$ is a boundary layer such as seen in Fig.~\ref{fig1}.
As horizontal components of the velocity grow at the expense of radial components (baryon conservation), the boundary layer develops  high shear, and is subject to KH instability \citep{drazin}. 

\item{Composition gradient.} Between $r_3$ and $r_4$ is the region that can develop a composition gradient, perhaps maintained at the margin of KH instability \citep{drazin}. This is not part of the original MLT \citep{bv58}; 
it is the {\em composition} problem of \S\ref{s-alvio}. A ``fix'' must be added (e.g., ``semi-convection" and/or  ``overshoot'' regions). This has led to a search for algorithms to extend MLT, whose justification is that they are at least empirically desirable if not always self-consistent. 

\item{Radiative.} Beyond $r_4$ is the ``radiative'' region of MLT. Unlike MLT, waves are automatically implied by flexing of the boundary layers ($r_1$ to $r_4$).
\end{enumerate}

This crude cartoon is to be understood as a snapshot, which flexes and bends;  its average properties are to be identified with stellar properties. The layers between $r_1$ and $r_4$ may be relatively thin in radial extent (see Fig.~\ref{figo16}).
Regions 2 and 3 are not defined within MLT; they are part of the {\em ends} problem of \S\ref{s-alvio}.  In part
region 4 is not defined within MLT because MLT assumes homogeneous composition \citep{bv58}; in MLT additional physics must be assumed concerning composition gradients to bridge this gap.  This region is often treated as ``semi-convective'', or partially mixed by  ``overshoot''. 
Regions 2, 3, 4, and 5 all support waves and relate to the {\em wind and waterline} problem (\S\ref{s-alvio}). Any improvement over MLT will certainly affect regions 2, 3 and 4.

\subsection{Composition profile}
\begin{figure*}[h]
\figurenum{4}
\label{figo16}
\includegraphics[angle=0,scale=0.38]{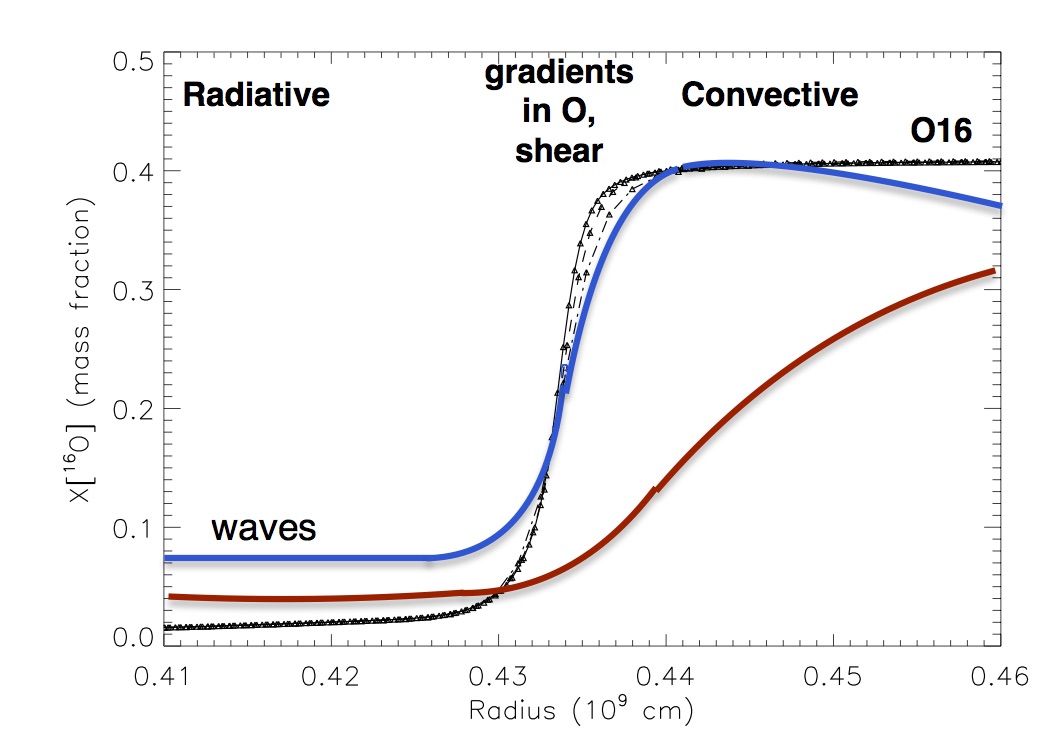}

\caption{Comparison of the $^{16}$O composition profiles (black), averaged over angle, for different resolutions (``med-res'' ($384 \times 256^2$, dots) and ``hi-res'' cases ($768\times 512^2$, heavy dots)  of \citep{viallet2013}, and a ``very-hi-res'' case ($1536\times1024^2$, heavier dots, the ``Perth'' simulation, see \cite{321D}), at the bottom boundary. This shape has striking similarity to that inferred from asteroseismology \citep{ehsan} for convective cores of red giants. See also discussion of the C+C shell in \cite{andrea,andrea2}. Overdrawn are the rms velocity profiles averaged over a spherical shell; red is radial  and blue is  for the two horizontal directions. Instantaneous profiles are not exactly spherical or constant but dynamic.
The velocities do not go to zero  outside the convective zone ($r<0.43 \times 10^9$cm ) because of significant but small wave motion (see Fig~\ref{fig1}).
}
\end{figure*}

Fig.~\ref{figo16} shows a composition profile of $^{16}$O at the lower boundary of the oxygen burning shell. To the extent that they are resolved the 3D ILES simulations automatically produce dynamically self-consistent boundary physics for turbulent flow. 

The rms velocity profiles are shown for radial (red) and horizontal (blue) motion. The dominance of horizontal velocity (blue) for $r < 0.43 \times 10^9$\,cm indicates g-mode waves. The boundary dynamics are those discussed in \cite{321D}. The need for a turning flow requires that the radial velocity (red) approach zero faster than the horizontal velocities (blue), which have a peak near the boundary.  There is a well-mixed region in  which this braking and turning occur, and which is sub-adiabatic.
Here the Brunt-V\"ais\"al\"a frequency is real; waves are generated by convective fluctuations and propagate from this region \citep[see \S\ref{Swavegen}, and Fig.~14,][]{woodward2015}. 

\placefigure{4}

The asymmetric behavior of convective boundaries, similar to that discussed in
\cite{gabriel,montalban}, can naturally arise in this braking region ($0.43 \times 10^9 < r < 0.44 \times 10^9$ cm in Fig.~\ref{figo16}). It is a consequence of carefully joining turbulent flow to a radiative region.

The sigmoid shape of the boundary is strikingly similar to that of the averaged mean flow of \cite{garaud1}, obtained with DNS for a shear flow setup;  see their Figure 3 (our use of the bottom boundary requires a flip in orientation). \cite{321D} showed that a turning plume would have such a shearing interface, even without differential rotation. The boundary conditions for differential rotation and convection seem to have a deep family resemblance.

\cite{andrea,andrea2} obtain a similar ``sigmoid-like'' profile for a carbon-burning shell. This shape 
from simulations has striking similarity to that inferred from astero-seismology \citep{ehsan} for the outer boundaries of convective cores of red giants. 
At first sight this is puzzling; radiative diffusion has a very minor  role in carbon and oxygen burning, but is more important in red giants \citep{viallet2013}. It suggests that {\em the shape of the composition gradient may be insensitive to the effects of radiative diffusion.} A clue appears in Fig.~\ref{figo16}, which shows a significant shear due to the horizontal velocities (blue curve), which may drive Kelvin-Helmholtz (KH) instabilities \citep{321D}. In the low Mach-number limit, this is a kinetic instability, driven just by the shape of the horizontal velocity in the radial direction (\cite{drazin}, see Fig.~8.3), and would be insensitive to radiative diffusion.
See  \S3.3 in \cite{woodward2015} for further discussion.

Consider a global perspective: as nuclear burning  proceeds the composition gradient steepens until KH instabilities induce some mixing to flatten it again.
The gradient in velocity profile evolves toward neutral stability to KH, and lies near the composition profile that it sculpts, as shown in
Fig.~\ref{figo16}.

An increase in entropy inside the convective zone tends to increase the size of the turbulent region. However, the amount of increase may be sensitively dependent upon the gradients of thermal and compositional entropy in the boundary regions, as well as the turbulent velocity.
A decrease in entropy gives a tendency to shed turbulent layers, which dissipate when no longer driven. Again there may be a  sensitive dependence upon the gradients of thermal and compositional entropy in the boundary regions, as well as the turbulent velocity.
This is essentially the Richardson criterion for stability for a layered system; \citep[][ \S10.2.3]{turner},
 
But what determines the velocity field at the interface? There is a similarity to the previous discussion (\S\ref{Smethods}) of the turbulent cascade and ILES. Approaching the convective boundary resembles approaching the Kolmogorov scale in velocity space; near the boundary only the small scales can ``fit''. 
Turbulence implies a shear layer at the interface.

\section{Conclusion}\label{Conclusion}

We have systematically re-examined the unphysical aspects of MLT in 1D stellar evolution, which were summarized in \citep{alvio87}. We have (1) performed 3D turbulent implicit large eddy simulations \citep[ILES,][]{iles07} of the classical B\"ohm-Vitense problem of stellar convection \citep{bv58}, (2) applied Reynolds-averaging (RA) to these numerical results, and (3) developed several analytic approximations to illustrate the physics involved. The RA-ILES combination is "exact" to the word length of the computer used, unlike the Reynolds-averaged Navier-Stokes equations (RANS) which are not closed \citep{tritton}. This advantage is a consequence of the sub-grid behavior of RA-ILES, which mimics a Kolmogorov turbulent cascade, allowing both stellar size scales and turbulent scales to be described.

The evolution of the turbulent kinetic energy is chaotic, involving a few dominant modes \citep[unstable 3D rolls,][]{lorenz}.
Driving and damping of turbulent convection are out of phase; this makes interpretation of time aspects of numerical convergence a challenge, but provides insights into the dynamics of turbulence and stellar variability.
We show an example of such turbulent pulses which are not significantly damped in 30 turnover times;
stellar convection can attain a {\em dynamic} steady-state, with chaotic fluctuations.

The RA-ILES simulations automatically provide the velocity structure at boundaries, including composition profile,
structure of deceleration region, and the basis for prediction of wave generation.
ILES allows us to describe a turbulent boundary layer, which uses the KH instability to sculpt a composition profile
similar to that inferred from astereoseismology \citep{ehsan}. Entrainment of stable matter occurs at these dynamic boundary layers \citep{ma07b}.
The process of turning the flow at the boundary also insures that waves are generated, connecting the convective motion to the wave generation.

We will quantify the resolution errors due to zoning in the companion paper \citep{wda2}, and predict the turbulent dissipation length (the `mixing length') from the simulations.

\acknowledgements
We thank Dr. Maxime Viallet and Dr. Miro Moc\'ak, whose work was an important foundation for this paper.
We thank the Theoretical Astrophysics Program (TAP) at the University of Arizona, and Steward Observatory for support.
This work was supported by resources provided by the Pawsey Supercomputing Centre with funding from the Australian Government and the Government of Western Australia.
This work used the Extreme Science and Engineering Discovery Environment (XSEDE), which is supported by National Science Foundation grant number OCI-1053575.  Some computations in our work made use of ORNL/Kraken and TACC/Stampede.

AC acknowledges partial support from NASA Grant NNX16AB25G. AC acknowledges the use of resources from the National Energy Research Scientific Computing Center (NERSC), which is supported by the Office of Science of the U.S. Department of Energy under Contract No. DE­AC02­05CH11231.

The authors acknowledge support from EU­FP7­ERC­2012­St Grant 306901. RH acknowledges support from the World Premier International Research Centre Initiative (WPI Initiative), MEXT, Japan. This article is based upon work from the ¤½ChETEC¤ COST Action (CA16117), supported by COST (European Cooperation in Science and Technology). CG, RH, and CM thank ISSI, Bern, for their support on organising meetings related to the content of this paper. CG acknowledge support from the Swiss National Science Foundation and from the Equal Opportunity Office of the University of Geneva.

This work used the DiRAC@Durham facility managed by the Institute for Computational Cosmology on behalf of the STFC DiRAC HPC Facility (www.dirac.ac.uk). The equipment was funded by BEIS capital funding via STFC capital grants ST/P002293/1 and ST/R002371/1, Durham University and STFC operations grant ST/R000832/1. This work also used the DiRAC Data Centric system at Durham University, operated by the Institute for Computational Cosmology on behalf of the STFC DiRAC HPC Facility. This equipment was funded by BIS National E­infrastructure capital grant ST/K00042X/1, STFC capital grants ST/H008519/1 and ST/K00087X/1, STFC DiRAC Operations grant ST/K003267/1 and Durham University. DiRAC is part of the National E­ Infrastructure.
We acknowledge PRACE for awarding us access to resource MareNostrum 4 based in Spain at Barcelona Supercomputing Center. The support of David
Vicente and Janko Strassburg from Barcelona Supercomputing Center, Spain, to the technical work is gratefully acknowledged.


\begin{thebibliography}

%
%
\bibitem[Acton(1970)]{acton} Acton, F. S., 1970, {\it Numerical Methods That Work}, Harper International Edition, Harper \& Row, New York
%
\bibitem[Aerts, et al.(2010)]{aerts} Aerts, C., Chistensen-Dalsgaard, J., \& Kurtz, D. W., 2010, Asteroseismology,
(Berlin: Springer)

\bibitem[Apsden, et al.(2008)]{apsden} Apsden, A. Nikiforakis, N., Dalziel,, S., \& Bell, J. B., 2008, Comm APP. Math. jand Compp. Sci., 3, 103

\bibitem[Arnett(1968)]{wda68} Arnett, W. D., 1968, Nature, 219, 1344



 





\bibitem[Arnett(1996)]{wda96} Arnett, D., 1996, {\it Supernovae and
Nucleosynthesis}, Princeton University Press, Princeton NJ

\bibitem[Arnett(2015)]{arnett2015} Arnett, W. D., 2015, IAUS 307, 459
%
\bibitem[Arnett, Meakin \& Young(2009)]{amy09vel} Arnett, W. D., Meakin, C., \& Young, P. A., 2009, \apj, 690, 1715 
%
\bibitem[Arnett \& Meakin(2010)]{am10rot} Arnett,  W. D. \& Meakin, C., 2010, IAUS, 265, 106
%
\bibitem[Arnett, Meakin \& Young(2010)]{amy10sub} Arnett, W. D., Meakin, C., \& Young, P. A., 2010 , \apj, 710, 1619 
%
%
\bibitem[Arnett \& Meakin(2011)]{am11turb} Arnett, W D., \& Meakin, C., 2011, \apj, 741, 33
%

\bibitem[Arnett \& Meakin(2016)]{am16key} Arnett, W D., \& Meakin, C., 2016, Reports of Prog. Phys., 79, 2901

%
\bibitem[Arnett, et al.(2015)]{321D} Arnett, W. D., Meakin, C. A., Viallet, M., Campbell, S. W., Lattanzio, J. C. \& Moc\'ak, M., 2015, \apj, 809, 30 

\bibitem[Arnett \& Moravveji(2017)]{ehsan} Arnett, W. D., \& Moravveji, E., 2017, \apj, 836L, 19

\bibitem[Arnett, et al.(2019)]{wda2} W. D. Arnett, C. Meakin, R. Hirschi, A. Cristini, C. Georgy, S. Campbell, L. Scott, E. Kaiser, paper II, submitted

%
\bibitem[Asplund, et al.(2009)]{asplund} Asplund, M.,  Grevesse, N., Sayval, A. J., Scott, P., 2009,
\araa, 47, 481

\bibitem[Asplund, Ludwig, Nordlund \& Stein(2000)]{alns00} Asplund, M, Ludwig, H.-G., Nordlund, \AA,  Stein, R. F.. 2000, \aap, 359, 669 

\bibitem[Basu \& Antia(1997)]{basu97} Basu, S. \& Antia, H. .,  1997, \mnras 287, 189

%
%
%
%
%
\bibitem[B\"ohm-Vitense(1958)]{bv58} B\"ohm-Vitense, E., 1958, \zap,
46, 108
%
\bibitem[B\"ohm-Vitense(1989)]{bv89} B\"ohm-Vitense, E., 1989, Introduction to Stellar Astrophysics, Vol. 2,  Stellar Atmospheres, Cambridge University Press 

\bibitem[Bressan, et al.(2012)]{bressan} Bressan, A., et al., 2010, \mnras, 427, 127B

\bibitem[Buldgen, et al.(2017)]{buldgen} Buldgen, G., et al., 2017, \aap 
%

\bibitem[Canuto \& Mazzitelli(1991)]{cm91} Canuto, V. M. \& Mazzitelli, I., 1991, \apj, 370, 295


\bibitem[Canuto, Goldman \& Mazzitelli(1996)]{can96} Canuto, V. M., Goldman, I.,
\& Mazzitelli, I., \apj, 473, 550



%
\bibitem[Canuto(2011a)]{can11a} Canuto, V. M. 2012, \aap, 528, A76
%
\bibitem[Canuto(2011b)]{can11b} Canuto, V. M. 2012, \aap, 528, A77
%
\bibitem[Canuto(2011c)]{can11c} Canuto, V. M. 2012, \aap, 528, A78
%
\bibitem[Canuto(2011d)]{can11d} Canuto, V. M. 2012, \aap, 528, A79
%
\bibitem[Canuto(2011e)]{can11e} Canuto, V. M. 2012, \aap, 528, A80

\bibitem[Chandrasekhar(1939)]{chandra} Chandrasekhar, S., 1939, An Introduction to the study of Stellar Structure, University of Chicago Press, Chicago

%
%
%
\bibitem[Clayton(1968)]{ddc} Clayton, D. D., 1968, Principles of Stellar Evolution and Nucleosynthesis, McGraw-Hill, New York
%
%
%

\bibitem[Colella \& Woodward(1984)]{cw84} Colella, P., \& Woodward, P., 1984,\jcp, 54, 174


%
%

%
\bibitem[Cristini, et al.(2017)]{andrea} Cristini, A., et al., 2017, \mnras, 471, 279
%
%
\bibitem[Cristini, et al.(2018)]{andrea2} Cristini, A., et al., 2018, submitted.
%
\bibitem[Davidson(2001)]{davidsonmhd} Davidson, P. A., 2001, An Introduction to Magnetohydrodynamics, Cambridge University Press

\bibitem[Davidson(2004)]{davidson} Davidson, P. A., 2004, Turbulence, Oxford University Press

%
%
%

\bibitem[Drake(2006)]{drake} Drake, R. P., 2006, {\it  High-Energy-Density Physics}, Springer, Berlin

\bibitem[Drazin(2002)]{drazin} Drazin, P. G., 2002, {\it Introduction to Hydrodynamic Stability}, Cambridge University Press, Cambridge

\bibitem[Ekstr\"om, et al.(2012)]{ekstrom} Ekstr\"om, S., et al., 2012, \aap, 537, A146
%
%
%
\bibitem[Eggleton, Dearborn \& Lattanzio(2008)]{edl08} Eggleton, P., Dearborn, D., \& Lattanzio, J.,
 2008, \apj, 677, 581
 
%
%
%
%
\bibitem[Featherstone \& Hindman(2016)]{featherstone} Featherstone, N. A., \& Hindman, B. W., 2016, \apj, 830, 15
 
%
%
%
\bibitem[Frisch(1995)]{frisch} Frisch, U., 1995, {\it Turbulence},
Cambridge University Press, Cambridge
%
%
\bibitem[Gabriel, et al.(2014)]{gabriel} Gabriel, M., Noels, A., Montalb\'an, J., Miglio, A., 2014, \aap, 569, A63

\bibitem[Gabriel \& Belkacem(2018)]{gabriel18} Gabriel, M. \& Belkacem, K., 2018, \aap, submitted

\bibitem[Gagnier \& Garaud(2018)]{gagnier-garaud} Gagnier, D. \& Garaud, P., 2018, submitted

\bibitem[Garaud, Gagnier \& Verhoeven(2017)]{garaud1} Garaud, p., Gagnier, D., \& Verhoeven, J.,, 2017, \apj, 837, 133

%

\bibitem[Gough(1968)]{gough} Gough, D. O., 1968, \aj, 72, 799
%
\bibitem[Gough(1977)]{gough77} Gough, D. O., 1977, in {\it Problems of stellar convection}, Proc. 38th Coloquium, Nice, France, Aug. 1976, (A78-28526 11-90) Berlin and New York, Springer-Verlag, 1977, p. 15-56

\bibitem[Grinstein, Magolin \& Rider(2007)]{iles07} Grinstein, Magolin, \& Rider, 2007, Implicit Large Eddy Simulations, Cambridge University Press

\bibitem[Hansen \& Kawaler(1994)]{hansenkawaler} Hansen, C. J., \& Kawaler, S., 1994, {\it Stellar Interiors: Physical Principles, Structure, and Evolution}, 1st. ed.,  Springer-Verlag, New York
%

\bibitem[Hansen, Kawaler, \& Trimble(2004)]{hansenkawaler2} Hansen, C. J., Kawaler, S., \& Trimble 2004, {\it Stellar Interiors: Physical Principles, Structure, and Evolution}, 2nd. ed.,  Springer-Verlag, New York


\bibitem[Hanasoge et. al(2010)]{hanasoge-apj}Hanasoge, S. M., Duvall, T. L.., \& DeRosa, M. L., 2010, \apj, 712, L98

\bibitem[Hanasoge et. al(2012)]{hanasoge-nas}Hanasoge, S. M., Duvall, T. L.., \& Sreenivasan, K. R., 2012, Proceed. lN. A. S., 109, 11928

\bibitem[Hanasoge \& Sreenivasan(2014)]{hanasoge} Hanasoge, S. M., \& Sreenivasan, K. R., 2014, Solar Physics, 289, 3403
 
%
\bibitem[Holmes, Lumley \& Berkooz(1996)]{holmes} Holmes,P., Lumley, J., \& Berkooz, G., 1996, Turbulence, Coherent Structures, Dynamical Systems and Symmetry, Cambridge University Press

\bibitem[Jones, et al.,(2017)]{jones17} Jones, S., R. Andrassy, S. Sandalski, A. Davis, P. Wood-
ward and F. Herwig, 2017, \mnras, 465, 2991

\bibitem[Joyce \& Chaboyer(2018)]{joyce-chaboyer} Joyce, M., \& Chaboyer, B., 2018, \apj, 856, 10

\bibitem[Kippenhahn \& Weigert(1990)]{kippen} Kippenhahn, R. \& Weigert, A. 
1990, {\it Stellar Structure and Evolution}, Springer-Verlag


%
%
\bibitem[Kolmogorov(1941)]{kolmg41} Kolmogorov, A. N., 1941, Dokl. 
Akad. Nauk SSSR, 30, 299
%
\bibitem[Kolmogorov(1962)]{kolmg} Kolmogorov, A. N.,1962, J. Fluid 
Mech., 13, 82
%
%
\bibitem[Landau \& Lifshitz(1959)]{llfm} Landau, L. D. \& Lifshitz, E. M., 1959, Fluid Mechanics,
Pergamon Press, London. 
%


%
\bibitem[Leveque(2002)]{leveque} Leveque, R. J., 2002, Finite Volume Methods for Hyperbolic Problems, Cambridge University Press, Cambridge,  UK

\bibitem[Lighthill(1978)]{lighthill} Lighthill, J., 1978, Waves in Fluids, Cambridge University Press, Cambridge, UK

%
%

\bibitem[Lorenz(1963)]{lorenz} Lorenz, E. N., 1963, Journal of Atmospheric
Sciences, 20, 130

\bibitem[Ludwig, Freytag, \& Steffen(1999)]{lfs99} Ludwig, H.-G., Fryetag, B., \& Steffen, M., 1999, \aap, 346, 111
%
%
\bibitem[Maeder(1999)]{maeder} Maeder, A., 1999, {\it Physics, Formation and Evolution of Rotating Stars},
Springer, Berlin
%
%
%
\bibitem[Manneville(2010)]{manneville} Manneville, P., 2010, {\it Instabilities, Chaos, and Turbulence}, Imperial College Press, London
%
%
%
%
\bibitem[Meakin \& Arnett(2007b)]{ma07b} Meakin, C. A., \& Arnett, W. D., \apj, 667, 448
%
\bibitem[Meakin \& Arnett(2010)]{ma10apss} Meakin, C. A., \& Arnett, W. D., \apss, 328, 22

\bibitem[Miesch, et al.(2012)]{miesch} Miesch, M., Featherstone, N., Rempel, M., \& Trampedach, R., 2012, \apj, 757, 128

%
%
%
%
%
%
\bibitem[Moc\'ak, et al.(2014)]{miro2014} Moc\'ak, M., Meakin, C., Viallet, M., \& Arnett, D., 2014, arXiv, 1401.5176 

\bibitem[Moc\'ak, et al.(2018)]{miro} Moc\'ak, M., Meakin, C., Arnett, D., \& Campbell, S., 2018, submitted.

\bibitem[Montalb\'an, et al.(2013)]{montalban} Montalb\'an, J., Miglio, A., Noels, A., et al. 2013, \apj, 766, 118

%
\bibitem[Nordlund, Stein, \& Asplund(2009)]{nsa} Nordlund, A., Stein, R., \& Asplund, M., 2009,
\url{http://ww.livingreviews.org/lrsp-2009-2}
%
%

\bibitem[O'Mara, et al.(2018)]{omara} O'Mara, B., Miesch, M. S., Featherstone, N. A., Agustson, K. C., 2018, arXiv:1603.06107v1,
Advances in Space Research, submitted

\bibitem[Orvendahl, et al.(2018)]{orvendahl} Orvendahl, R., Calkins, M., Featherstone, N., \& Hindman, B., 2018, \apj, submitted

\bibitem[Parker(1979)]{parker} Parker, E. N., 1979, Cosmic Magnetic Fields,Clarendon Press, Oxford
%
%
%
%
%

\bibitem[Pope(2000)]{pope} Pope, S. B., 2000, {\it Turbulent Flows}, 
Cambridge University Press, Cambridge, GB

\bibitem[Porter \& Woodward(2000)]{porterwoodward} Porter, D. H. and P. R. Woodward. 2000, \apjs 127, 159


\bibitem[Radice, Couch \&  Ott(2015)]{radice15} Radice, D., Couch, S., \& Ott, C., 2015, Computational Astrophysics and Cosmology, 2, 7 (arXiv:1501.03169)

\bibitem[Renzini(1987)]{alvio87} Renzini, A., 1987, \aap, 188, 49
%
%

%
%
\bibitem[Smith \& Arnett(2014)]{nathan2014} Smith, Nathan, \& Arnett, W. D., 2014, \apj, 785, 82
%
%
%
%
\bibitem[Stein \& Nordlund(1989)]{sn89} Stein, R. F., \& Nordlund, A., 1989, \apj, 342, 95
%
%
\bibitem[Stein \& Nordlund(1998)]{sn98} Stein, R. F., \& Nordlund, A., 1998, \apj, 499, 914
%

\bibitem[Sytine, et al.(2000)]{sytine} Sytine, I. V., Porter, D. H., Woodward, P. R., Hodson, S. W., Winkler, K.-H., Journal of Computational Physics 158 (2000), 225
%
%
\bibitem[Tennekes \& Lumley(1972)]{tennekes} Tennekes, H., \& Lumley, 
J. L., 1972, {\it A First Course in Turbulence}, MIT Press, Cambridge MA
%
\bibitem[Trampedach \& Stein(2011)]{tramp-stein} Trampedach, R., \& Stein, R.. F., \apj, 731, 78 %

\bibitem[Trampedach, et al.(2014)]{tramp} Trampedach, R., Stein, R.. F., Christensen-Dalsgaard, J., Nordlund, A., Asplund, M.  2014, \mnras, 

%
\bibitem[Tritton(1988)]{tritton} Tritton, D. J., {\it Physical Fluid
Dynamics}, 2nd ed., Oxford University Press, Oxford UK
%



\bibitem[Turner(1973)]{turner} Turner, J. S. 1973, Buoyancy Effects in Fluids, Cambridge University Press, Cambridge, UK

\bibitem[Tzeferacos, et al.(2018)]{dqlB} Tzeferacos, P. et al., 2018, Nature Communications,  DOI: 10.1038/s41467-018-02953-2
%
\bibitem[Unno, et al.,(1989)]{unno}Unno, W., Osaki, Y., Ando, H., Saio, H., \& Shibahashi, H. 1989, Nonradial Oscillations of Stars (2nd ed.; Tokyo: Univ. of Tokyo Press)


\bibitem[Viallet, et al.(2011)]{viallet2011} Viallet, M., Baraffe, I., \& Walder, R., 2011,
\aap, 531, 86
%
\bibitem[Viallet, et al.(2013)]{viallet2013} Viallet, M., Meakin, C., Arnett, D., Moc\'ak, M., 2013, \apj, 769, 1
%

\bibitem[Vinyoles, et al.(2017)]{serenelli} Vinyoles,, N., Serenellii, A., et. al., 2017, \apj,  835, 202

\bibitem[Vitense(1953)]{vitense53} Vitense, E., 1953, \zap, 32, 135
%


\bibitem[Warhaft(2002)]{warhaft} Warhaft, Z., 2002, PNAS 99, 2481 

\bibitem[Woodward(2007)]{woodward} Woodward, P. R., 2007, in {\it Implicit
Large Eddy Simulantions}, ed. F. F. Grinstein, L. G. Margolin, \& W. J.
Rider, Cambridge University Press, p. 130 
%
	
\bibitem[Woodward et al.(2015)]{woodward2015} Woodward, Paul R., Herwig, Falk, Lin, Pei-Hung,
2015, \apj, 798, 49





\end{thebibliography}
\end{document}